\documentstyle[titlepage,12pt]{article}
\newcommand{\beq}{\begin{equation}}
\newcommand{\eeq}{\end{equation}}
\newcommand{\eq}[1]{eq.(\ref{#1})}
\title {
\begin{flushright} PSU/TH/151
\end{flushright}
\bigskip\bigskip\bigskip
NEW CORRECTIONS OF ORDER $\alpha^6$ TO $S$-LEVELS OF TWO-BODY SYSTEMS
}
\author
{Michael I. Eides\thanks{E-mail address: eides@lnpi.spb.su}\\
Petersburg Nuclear Physics Institute,\\
Gatchina, St.Petersburg 188350, Russia\\
\and \medskip
and\\
Howard Grotch\thanks{E-mail address: h1g@psuvm.psu.edu}\\
Department of Physics, Pennsylvania State University,\\
University Park, Pennsylvania 16802, USA\medskip}
\date{December, 1994}

\begin{document}
\maketitle
\begin{abstract}
New corrections to the energy of $S$-levels of positronium of order
$m\alpha^6$ which are as large as several hundred kilohertz are obtained.
A new recoil correction of order $\alpha(Z\alpha)^5(m/M)m$ to the Lamb shift
in hydrogen is calculated. This correction turns out to be too small from
the phenomenological point of view.
\end{abstract}

{\bf 1}. Recent progress in the spectroscopy of positronium
\cite{fee,chu,danz,hat,cont} triggered theoretical work on the corrections
of order $\alpha^6m$ to the positronium energy levels. All logarithmic
corrections of this order to $S$-levels were calculated recently in
\cite{fell,kmy1}. Complete results for the corrections
of order $\alpha^6m$ to $P$-levels were obtained in \cite{kmy2}. As
emphasized in this last work the large magnitude of the nonlogarithmic
corrections to $P$-levels suggest that calculation of corresponding
nonlogarithmic corrections to $S$-levels is also important. Some of these
corrections are already known, e.g., contributions induced by the two- and
three-photon annihilation kernels \cite{abz,dr,aab}. We present below
results of the calculation of nonlogarithmic contributions of order
$\alpha^6m$ to the $S$-levels of positronium induced by radiative
corrections to the Breit potential and  by the polarization insertions in
the graphs with two-photon exchange.

A new radiative-recoil correction of order $\alpha(Z\alpha)^5(m/M)m$ to the
Lamb shift in hydrogen induced by a polarization operator insertion in the
two-photon exchange graph is also calculated in this note. Recent
experimental achievements in measuring $1S-2S$ splitting in hydrogen
\cite{han} and the well-known results on the $2S$ Lamb shift \cite{lp,ps,hp}
clearly demonstrate that theoretical calculation of all corrections to the
Lamb shift of the order of several kHz for the $1S$-state and about $1$ kHz
for the $2S$-state is necessary. Several such contributions were obtained
quite recently \cite{gp,kar,gkme} and the result presented below is one more
such contribution (for more detailed description of the current theoretical
status of the Lamb shift calculations see, e.g.  \cite{gr}).

{\bf 2}. Let us consider first corrections of order $\alpha^6m$ to the
$S$-levels of positronium connected with radiative insertions in the graph
with one-photon exchange in Fig 1. As is well known this graph leads to the
Breit potential. One may easily obtain the radiatively corrected expression
for the Breit potential in the form\footnote{The annihilation diagram
contribution is missing in this expression since we do not consider
annihilation contributions in this paper.} (see, e.g. \cite{gh} and paper in
preparation)

\beq \label{breit}
U({\bf p},{\bf r})=-\alpha\{\frac{1}{r}
-\pi\frac{1+8f'_1+2f_2}{m^2c^2}\delta^3({\bf r})
+\frac{4\pi p}{m^2c^2}\delta^3({\bf r})
\eeq
\[
+\frac{{\bf r}({\bf r}{\bf p}){\bf p}}{2m^2c^2r^3}+\frac{{\bf p}^2}{2m^2c^2r}
-(3+4f_2)\frac{{\bf s}{\bf l}}{2m^2c^2r^3}
\]
\[
+\frac{(1+f_2)^2}{4m^2c^2}\{s_i,s_j\}(\frac{\delta_{ij}}{r^3}
-3\frac{r_ir_j}{r^5})
-\frac{(1+f_2)^2\pi}{m^2c^2}(\frac{4}{3}{\bf s}^2-2)\delta^3({\bf
r})\},
\]

where $m$ is the electron mass, $\bf p$ is the relative momentum of the
electron and positron, $\bf r$ is their relative position, $f_{1}'$ is the
slope of the Dirac formfactor, $f_2$ is the Pauli formfactor at zero
momentum transfer and $p$ is the polarization operator contribution. With
two-loop accuracy we have

\beq
f'_1=\frac{e_1}{m^2}\frac{\alpha}{\pi}+\frac{e_2}{m^2}(\frac{\alpha}{\pi})^2,
\eeq
\[
f_2=g_1\frac{\alpha}{\pi}+g_2(\frac{\alpha}{\pi})^2,
\]
\[
p=p_1\frac{\alpha}{\pi}+p_2(\frac{\alpha}{\pi})^2.
\]

It is an easy task now to obtain corrections of order $\alpha^6m$ to
the positronium energy levels

\beq
\Delta E_{F_1}=e_2\frac{\alpha^6}{\pi^2n^3}m\delta_{l0}
=0.469~94~\frac{\alpha^6}{\pi^2n^3}m\delta_{l0},
\eeq
\[
\Delta E_{F_2,|l=0}=g_2\frac{\alpha^4m}{4n^3}
=-0.082\frac{\alpha^6m}{\pi^2n^3},
\]
\[
\Delta E_{p2}=-p_2\frac{\alpha^6}{2\pi^2n^3}m\delta_{l0}
=-\frac{41}{324}\frac{\alpha^6}{\pi^2n^3}m\delta_{l0},
\]

where we used the value of the two-loop contribution $e_2$ to the slope of
the Dirac formfactor obtained numerically in \cite{ab} and analytically in
\cite{bmr}, the explicit results for the two-loop electron magnetic moment
$g_2$ \cite{p,s}, and the two-loop irreducible vacuum polarization operator
\cite{ks} obtained a long time ago.

With the help of the effective potential in \eq{breit} we may also easily
calculate radiative corrections to the levels of positronium which have
nonvanishing angular momentum. Our results in this case reproduce and
confirm the respective results in \cite{kmy2,gkme}.

{\bf 3}.  Consider now corrections of relative order $\alpha^6$ to the
energy levels of two-body systems which are generated by the diagrams with
intermediate momenta which are high on the scale of the typical atomic
momenta. It is well known that all such corrections are generated by the
diagrams with two exchanged photons containing also either a polarization
operator insertion in one of the exchanged photons or radiative photon
insertions in the electron line (see, e.g. \cite{eks}). To sufficient
accuracy external electron lines in the diagrams under consideration may be
safely taken to be on-mass shell. It is not difficult to obtain an explicit
expression for the infrared divergent skeleton integral corresponding to the
sum of ladder and crossed diagrams in Fig.2. Direct integration over loop
momentum advocated in \cite{gy} leads to the following expression for the
skeleton integral

\beq          \label{genform}
\Delta E_{skel}=-32mM(Z\alpha)^2|\psi(0)|^2
\eeq
\[
\int_0^\infty\frac{dk}{\pi k}\int_0^\pi d\theta
\frac{\sin^2\theta(1+2\cos^4\theta)}{(k^2+4m^2\cos^2\theta)
(k^2+4M^2\cos^2\theta)}
\]
\[
=-16mM(Z\alpha)^2|\psi(0)|^2\int_0^\infty\frac{dk}{k^3}\frac{1}{(m^2-M^2)}
\{m\sqrt{1+\frac{k^2}{4m^2}}(\frac{1}{k}+\frac{k^3}{8m^4})
\]
\[
-M\sqrt{1+\frac{k^2}{4M^2}}(\frac{1}{k}+\frac{k^3}{8M^4})
-\frac{k^2}{8m^2}(1+\frac{k^2}{2m^2})
+\frac{k^2}{8M^2}(1+\frac{k^2}{2M^2})\},
\]

where $m$ and $M$ are the masses of the negatively and positively charged
particles, respectively, $Z$ is the charge of the positive particle in terms
of the proton charge and $\psi(0)$ is the value of the reduced mass
Schr\"odinger-Coulomb wave function at the origin.

All contributions to hydrogen Lamb shift of order $\alpha(Z\alpha)^5m$, both
recoil and nonrecoil, calculated over years by different methods
\cite{kks,bbf,bg}, may be obtained from the expression for the
skeleton integral in \eq{genform} by insertion of radiative corrections.

Consider first recoil contributions of order $\alpha(Z\alpha)^5m$ to the
Lamb shift in hydrogen. The contribution induced by the radiative
photon insertions in the electron line was obtained in \cite{bg}. With the
help of explicit expression in \eq{genform} above, it is easy to confirm the
result of \cite{dgo} that correction induced by the radiative photon
insertions in the heavy line is suppressed by the factor $(m/M)^2$ relative
to the contribution induced by the radiative photon insertion in the
electron line. It is also easy to see that the recoil correction
corresponding to the polarization operator insertion in the exchanged photon
is suppressed by the factor $m/M$ relative to respective nonrecoil
correction. Let us calculate this last correction. The general expression in
\eq{genform} contains the skeleton integral both for recoil and nonrecoil
corrections. The skeleton integral for the recoil corrections may be
obtained by subtracting the heavy pole residue in \eq{genform} and has the
form

\beq                       \label{skelrec}
\Delta E_{skel-rec}
=\frac{16(Z\alpha)^2|\psi(0)|^2}{m^2(1-\mu^2)}\int_0^\infty
\frac{kdk}{(k^2+\lambda^2)^2}
\{\mu\sqrt{1+\frac{k^2}{4}}(\frac{1}{k}+\frac{k^3}{8})
\eeq
\[
-\sqrt{1+\frac{\mu^2k^2}{4}}(\frac{1}{k}+\frac{\mu^4k^3}{8})
-\frac{\mu k^2}{8}(1+\frac{k^2}{2})
+\frac{\mu^3k^2}{8}(1+\frac{\mu^2k^2}{2})
+\frac{1}{k}\},
\]

where $\mu=m/M$, $\lambda$ is an auxiliary mass of the exchanged photon
which is omitted below,  and we transferred to a dimensionless integration
momentum measured in units of the electron mass.

For calculation of the radiative-recoil contribution to the
Lamb shift induced by the polarization operator insertions one has to make
a substitution in the integrand in \eq{skelrec}

\beq         \label{sub}
\frac{1}{k^2}\rightarrow \frac{\alpha}{\pi}I_{1}(k),
\eeq

where

\beq
{I_1(k)}= \int_0^1 dv \frac{v^2(1-v^2/3)}{4+(1-v^2)k^2}\:.
\eeq

However, the skeleton integrand in \eq{skelrec} behaves as $\mu/k^4$ at small
momenta and naive substitution in \eq{sub} leads to divergence. This
divergence $dk/k^2$ actually diminishes the power of the $Z\alpha$ factor
and the respective contribution turns out to be of order
$\alpha(Z\alpha)^4$. In order to get the recoil correction of order
$\alpha(Z\alpha)^5m$ we have to subtract the leading low
frequency asymptote of the product of the skeleton integrand and the
polarization operator. Then we obtain the integral for the radiative-recoil
correction (one has to insert an additional factor of $2$ which takes into
account possible insertions of the polarization in both photon lines)

\beq
\Delta
E_{r}=\frac{32(Z\alpha)^2|\psi(0)|^2}{m^2(1-\mu^2)}(\frac{\alpha}{\pi})
\int_0^\infty
\frac{dk}{k}\{I_{1}(k)
\left[\mu\sqrt{1+\frac{k^2}{4}}(\frac{1}{k}+\frac{k^3}{8})
\right.
\eeq
\[
\left.
-\sqrt{1+\frac{\mu^2k^2}{4}}(\frac{1}{k}+\frac{\mu^4k^3}{8})
-\frac{\mu k^2}{8}(1+\frac{k^2}{2})
+\frac{\mu^3k^2}{8}(1+\frac{\mu^2k^2}{2})+\frac{1}{k}\right]-\frac{\mu}{15k}\}.
\]

This integral contains also some contributions of higher order in
the electron-proton mass ratio and may be easily calculated numerically.
However, these higher order contributions are clearly negligible and we omit
them. Then we obtain the analytic result

\beq
\Delta E_{r}=(\frac{2\pi^2}{9}-\frac{70}{27})
\mu\frac{\alpha(Z\alpha)^5m}{\pi^2n^3}
(\frac{m_r}{m})^3.
\eeq

{\bf 4}. Consider now the contribution of order $\alpha^6$ induced by
insertion of the one-loop polarization operator for the positronium case.
Calculation is similar to the one for hydrogen. The analog of the skeleton
integral in \eq{genform}, for the case of equal masses, has the form

\beq           \label{posskel}
\Delta E=-16m^2(Z\alpha)^2|\psi(0)|^2\int_0^\infty\frac{dk}{k^4}
\{\frac{1}{\sqrt{k^2+4m^2}}
\eeq
\[
+\frac{k^3}{8m^4}-\frac{k^4(k^2+3m^2)}{8m^6\sqrt{k^2+4m^2}}
+\frac{k^5}{8m^6}\}
\]
\[
=-\frac{2\alpha^5m^5}{\pi
n^3}\int_0^\infty\frac{dk}{k^4}\{\frac{1}{\sqrt{k^2+4m^2}}
\]
\[
+\frac{k^3}{8m^4}-\frac{k^4(k^2+3m^2)}{8m^6\sqrt{k^2+4m^2}}
+\frac{k^5}{8m^6}\}.
\]

Consideration of the hydrogen case above teaches us that the integrand for
the radiative correction is given by the subtracted product of the vacuum
polarization operator and the skeleton integrand. Hence, contribution to the
energy levels of positronium of order $m\alpha^6$ is given by the expression
(remember combinatorial factor $2$)

\beq
\Delta E_{p1}=-\frac{4\alpha^6m^5}{\pi^2
n^3}\int_0^\infty\frac{dk}{k^2}\{I_{1}(k)\left[\frac{1}{\sqrt{k^2+4m^2}}
+\frac{k^3}{8m^4}-\frac{k^4(k^2+3m^2)}{8m^6\sqrt{k^2+4m^2}}
\right.
\eeq
\[
\left.
+\frac{k^5}{8m^6}\right]
-\frac{1}{30}\}=(\frac{\pi^2}{36}-\frac{5}{27})\frac{m\alpha^6}{\pi^2n^3}.
\]

{\bf 5}. Numerical values of the corrections obtained above are presented in
the Table.

\begin{center}
\begin{tabular}{|l|rl|r|r|}    \hline
          &                                          &  &$2S$ & $1S$
\\
\ $\Delta E$  & $$ &  &kHz &  kHz
\\ \hline
Positronium, $\Delta E_{F_1}$  &   $0.469\frac{\alpha^6}{\pi^2n^3}m$    & &
$111.05$ & $888.40$
\\ \hline
Positronium, $\Delta E_{F_2}$         &
$-0.082\frac{\alpha^6}{\pi^2n^3}m$      & & $-19.38$  &  $-155.02$
\\ \hline
Positronium, $\Delta E_{p2}$ &
$-\frac{41}{324}\frac{\alpha^6}{\pi^2n^3}m$ & & $-29.90$ & $-239.22$
\\ \hline
Positronium, $\Delta E_{p1}$ &
$(\frac{\pi^2}{36}-\frac{5}{27})\frac{\alpha^6}{\pi^2n^3}m$ & & $21.02$ &
$168.19$
\\ \hline
Hydrogen, $\Delta E_{r}$ &
$(\frac{2\pi^2}{9}-\frac{70}{27})
\frac{\alpha(Z\alpha)^5}{\pi^2n^3}\frac{m}{M}m$ & & $-0.05$ & $-0.41$
\\ \hline
\end{tabular}
\end{center}

In the case of positronium these corrections turn out to be of the same
order of magnitude as other corrections to the energy levels calculated
recently \cite{fell,kmy1,kmy2,abz,dr,aab} and are significant for
comparison of the theory with the current experimental results. In the
case of hydrogen the corrections for the $S$-levels obtained above are
about an order of magnitude smaller than the corrections of order
$\alpha^6(m/M)m$ for the $P$-levels obtained recently \cite{gkme} and are
too small to be interesting from the phenomenological point of view.
Detailed derivation of the results of this paper will be presented
elsewhere. In the case of positronium there remain some other yet unknown
contributions of order $\alpha^6m$ to the energy shift of $S$-levels. Work
on their calculation is in progress now.

\medskip
We are deeply grateful to V. A. Shelyuto and S. G. Karshenboim for
attracting our attention to inaccuracy of our subtraction procedure in
the preliminary version of this report. We are also deeply grateful to
I.B.Khriplovich for a helpful remark on the radiative corrections to the
Breit potential and to K. Pachucki for useful communications. This work was
done during the visit of M.E. to the Penn State University. He is deeply
grateful to the colleagues at the Physics Department of the Penn State
University for their kind hospitality.  This research was supported by the
National Science Foundation under grant \#NSF-PHY-9120102. Work of M.E.  was
also supported in part by grant \#R2E000 from the International Science
Foundation and by the Russian Foundation for Fundamental Research under
grant \#93-02-3853.

\newpage
\begin{center}\large Figure Captions
\end{center}
\vskip5cm

Fig.1.  One-photon exchange skeleton graph.\\

Fig.2. Two-photon exchange skeleton graphs.

\end{document}